\documentclass[10pt, a4paper, notitlepage]{article}

\usepackage{tgpagella}  
\usepackage{mathpazo}   

\usepackage{geometry}
\usepackage{amsmath}
\usepackage{graphicx}
\usepackage[numbers]{natbib}
\bibpunct{(}{)}{;}{a}{,}{,}

\usepackage[dvipsnames]{xcolor}
\usepackage[colorlinks=true,
  linkcolor=CadetBlue,
  citecolor=CadetBlue,
  urlcolor=CadetBlue]{hyperref}
\usepackage{url}

\usepackage{sectsty}
\sectionfont{\fontsize{12}{12}\selectfont}
       \subsectionfont{\fontsize{10}{10}\selectfont}

\newcommand{\change}[1]{\textcolor{black}{#1}}

\usepackage{fancyhdr} 
\fancypagestyle{firststyle}
{
   \fancyhf{}
   \fancyfoot[C]{\scriptsize{\textit{Minds and Machines}, published online in July 2021, pp.~1--18. This is the author's copy. The publisher's open access copy is available online at \url{https://doi.org/10.1007/s11023-021-09566-7}.}}
}

\begin{document}

\title{The Treachery of Images in the \\ Digital Sovereignty Debate}
\author{\normalfont Jukka Ruohonen \vspace{4pt}\\ University of Turku, Turku, Finland \\ \url{https://orcid.org/0000-0001-5147-3084}  \\ \texttt{juanruo@utu.fi}}

\maketitle

\begin{abstract}
This short theoretical \change{and argumentative} essay contributes to the
ongoing deliberation about the so-called digital sovereignty, as pursued
particularly in the European Union (EU). Drawing from classical political
science literature, the essay approaches the debate through paradoxes that arise
from applying classical notions of sovereignty to the digital domain. With these
paradoxes and a focus on the Peace of Westphalia in 1648, the essay develops a
viewpoint distinct from the conventional territorial notion of
sovereignty. Accordingly, the lesson from Westphalia has more to do with the
capacity of a state to govern. It is also this capacity that is argued to enable
the sovereignty of individuals within the digital realm.  With this viewpoint,
the essay further advances another, broader, and more pressing debate on
politics and democracy in the digital era.
\vspace{5pt}
\end{abstract}

\section*{Introduction}

\thispagestyle{firststyle} 

This theoretical \change{and argumentative} essay contributes to the debate on
so-called digital sovereignty. The term ``digital'' is used to narrow the scope;
the focus is on the Internet and particularly its so-called application layer at
which much of the action occurs. This layer includes the Web, electronic mail,
the domain name system, cloud computing, platforms, and so forth. The
alternative term ``cyber'' is wider, including the global positioning system and
space technologies, to name a couple of examples.

The core of the debate revolves around a question on whether and how the
``digital realm'' aligns with the ``physical realm''. No particular motivation
is required for an interlocutor. The question is intriguing for each and every;
robots, artificial intelligence, biomedical implants, virtual reality, and
whatever else comes to your mind. But as exciting as the philosophy of mind is,
it will sooner or later face the harsh realities of political philosophy. Within
these realities, the question is thorny already because all classical notions of
sovereignty belong to the physical realm. These notions are much older than
anything digital; sovereignty as a concept traces through the whole history of
Western philosophical thought, either explicitly or implicitly. Against this
backdrop, it is no real surprise that digital sovereignty has recently been
actively debated both in academia and in politics. In terms of academic
research, a number of recent surveys and essays already
exist~\citep{Christakis20, Couture19, Floridi20, Timmers19a}. Although these have
done good jobs at cataloging the terminology involved, raising important
arguments, and advancing the debate in general, they have often overlooked or
simplified some of the nuances, historical parallels, and political
underpinnings.

Regarding the political underpinnings, digital sovereignty has long been a baton
in geopolitics, with some countries using the concept in their political
rhetoric seeking justify increasing state control over the Internet. In a
similar vein, there has been a disparity between the political stances of the EU
and the United States; digital sovereignty has been advanced by the former,
whereas in the United States it has been viewed in a skeptical
light~\citep{Couture19}. Here, again, geopolitics are involved. In Europe the
debate aligns with the aspirations to gain technological independence and
resilience.

As the digital and physical realms increasingly intervene, not only are economic
interests present but the geopolitical dimension extends toward security and
defense, which, in particular, within the EU, are often debated by using a
related but still distinct concept of strategic autonomy. Given that sovereignty
alone is a conceptual bonanza---there are at least sixteen different notions of
the concept~\citep{Christakis20}, strategic autonomy can be reasonably excluded
from the paper's scope, although---and importantly, it can be also seen as a way
to achieve digital sovereignty~\citep{Timmers19b}. Recently, \change{for better
  or worse}, sovereignty as a concept seems to have also lost to strategic
autonomy particularly in the European security and defense
debates~\citep{Rahman21}. \change{In other words, at the moment---and like in
  the past, a fully sovereign European solution does not seem plausible in this
  policy space without the transatlantic link.} But even when excluding
strategic autonomy, more restrictions are needed to distinguish the paper from
the existing literature surveys \change{and essays}. What, then, to further
exclude or include?  For the paper's purposes, \change{a sensible} answer can be
found from the following simple quotation:

\begin{quote}
  ``\textit{Finally, the EU is pursuing a tech industrial policy under the strategically---and morally---ambiguous heading of `digital sovereignty.' Proponents of the concept toggle breezily between two definitions of 'sovereignty.' One is based on human-centered autonomy---each individual citizen is personally sovereign over their data, interactions with AI, etc. The other is a more Westphalian understanding of sovereignty: each state has an undisputed power monopoly within its borders.}'' \citep{Barker20}
\end{quote}

\change{When putting the political points aside---and also keeping in mind that
  the quotation supposedly reflects a sentiment held in some parts of the
  national security apparatus of the United States}, the quotation's value stems
from its pertinence to concisely frame the two core tenets in the digital
sovereignty debate. Thus, in what follows, the explicit framing contains
human-centered autonomy on the one hand and autonomy of states within their
territories on the other hand. \change{Within the polemical style}, there is
also \change{a further,} implicit framing in the quotation. This frame is
located in the claim that participants in the debate can ``\textit{toggle
  breezily}'' between the two parts in the explicit frame. As will be
elaborated, this alleged breeze leads to interesting theoretical
paradoxes. \change{For motivating} these paradoxes, one only needs to ask: how
can there be individual autonomy in the digital realm within which undisputed
power exists neither within nor between states?  The answer, as will be argued,
has to do with authority and power to which sovereignty is unavoidably
linked. Authority, in turn, requires a philosophical rather than a technological
perspective, although, as the essay proceeds, some technical examples must be
provided for illustrative purposes.

Even though the essay is theoretical, furthermore, some analytical tools are
still required. Like the debate in politics, the academic and engineering side
of the digital sovereignty debate can be read as a deliberation process. To read
such a process and to further participate in the deliberation, classical yet
analytical tools in political science provide a useful way
forward~\citep{Steiner08}. Thus, in what follows, a few of the main points in
the debate are recapitulated by trying to reflect upon the images others have
seen \change{about digital sovereignty}, and by trying to turn these reflections
into new \change{theoretical} images.

\change{But before} proceeding, a few passages must be written about the
analytical tools mentioned. After these \change{methodological} notes, the
paper's structure is simple: the two themes---the territorial sovereignty and
human-centered autonomy---are discussed in their consecutive
sections. \change{The former theme is discussed by focusing on classical
  definitions for state sovereignty, transborder access to digital assets and
  data protection laws, cyber war, and the role of large information technology
  companies. The Peace of Westphalia, the Holy Roman Empire, and the East India
  Company are used as historical parallels for elaborating the discussion. The
  second theme, in turn, is discussed by focusing on conflict theories and
  liberalism, following classical Western political philosophy. To address the
  paradoxes arising from the alignment of the two realms, international law,
  data protection, surveillance, large companies, and digital infrastructures
  are elaborated in conjunction with the political philosophy tenets. After
  these two sections, a few reflections and a concluding section follow.}

\section*{Miscomparisons}

What is a miscomparison? It is easy to say what it is not. It is not one of the
fallacies commonly appearing in debates. Nor is it one of the common vices of
argumentation~\citep{Aberdein16} to which so many weak minds are
vulnerable~\citep{Laroche18}. To give a practical example: during his heyday,
one famous Finnish populist pen introduced a number of brilliant comparative
sayings from his Moomin pencil box. For instance: whenever ``\textit{a melon and
  an apple each wear the same size baseball cap, everyone can see that just
  doesn't work}'' \citep{MarketWatch11}. Although no one still knows to whom or
to what the melon (Germany?) and the apple (Finland?) exactly referred, there is
a grain of wisdom in the quote, perhaps. Both melons and apples are delicious
fruits. Therefore, they should be comparable. However, the method of the
comparison, the baseball cap, is an awry technique for the task. Therefore,
``that just doesn't work''.

As it happens, this kind of a faulty reasoning is exactly what \citet{Sartori91}
famously discussed with his miscomparison framework involving dichotomous
concepts and their theoretical underpinnings. A classical example of such
concepts is the notorious friend-enemy framework of the notorious Carl Schmitt,
as also analyzed by \citet{Sartori89}. Further examples are not difficult to
imagine: true and false, zero and one, center and periphery, security and
insecurity, war and peace, and so forth and so on. In a series of papers
starting from the 1970s, \citeauthor{Sartori89} also developed various means to
identify miscomparisons often associated with such dichotomies. These have been
frequently used to discuss different comparison problems~(see, e.g.,
\citealt{Ahram11, Daly03}). Following \citeauthor{Sartori70}'s work
\citeyearpar{Sartori70, Sartori91}, four such means are typically discussed in
the literature.

The first is \textit{parochialism}: typically a case study that simply ignores
any categories postulated by general theories and comparative frameworks. When
further accompanied with custom ad-hoc terminology and general intellectual
laziness, a miscomparison often occurs in the form of a mislabel.  Mislabeling
is often accompanied with the second common error, \textit{misclassification},
typically caused by conceptual orderings that are derived from a single
criterion. If fruits are again classified with the baseball cap, a
fruit-fruit-fruit could result from classifying a melon, an apple, and a
snowberry; however, under a different classification scheme, a eatable-uneatable
could result as snowberries are poisonous to humans. In other words, there are
two problems here: choosing the reference baskets and putting the fruits into
the right baskets. As we shall see in a moment, there are many digital
sovereignty baskets, and it is not at all clear into which particular basket
which particular fruit should go. Sometimes, ``that just doesn't work''.

The third typical error, ''\textit{degreeism}'', stems from operationalization
of theoretical concepts and quantitative variables. Its essence is a belief that
continuous variables and continuum-based concepts are preferable over variables
or concepts with a dichotomous or an ordinal scale. There are situations for
which continuum-based concepts are well-suited. For instance, hybrid warfare may
blur the boundaries between military and civilian and covert and overt;
propaganda spread on a private platform may blur the boundary between real and
unreal; and so on~(cf.~\citealt{Chong14, Moskalenko17, Ruohonen20EJSR,
  Ruohonen21GOODIT}). In these cases it is not necessarily meaningful to operate
with strict theoretical dichotomies.  However, there are problems as well. As
there are seldom perfect things, all essentially dichotomous concepts become
imperfect without some arbitrary threshold turning the concepts back into their
dichotomous forms. \citet{Sartori91} asks: what is a ``80 percent democracy''?
The question is non-trivial. Clearly, there are no perfect democracies, but, at
some point, a threshold is required to distinguish democracies from other forms
of government.

The fourth error, or a problem, is \textit{conceptual stretching}, or, depending
on a situation, conceptual shrinking. In essence: whenever a concept is
stretched too much, it starts to mean everything and thus
nothing~\citep{Steiner08}. In other words, social sciences, in particular,
require broad universal concepts---public, private, military, civilian, war,
peace, democracy, justice, etc.---yet these frequently turn into amorphous
conceptualization that are difficult to operate with and to
measure~\citep{Sartori70}. But, here, please take a note: with this simple
statement, a lazy, undiscerning mind may have already committed itself to
parochialism; nowhere is it stated that a difficulty would imply an
impossibility (cf.~\citealt{WJP20}). Likewise, it should be stressed
  that conceptual stretching is often necessary when a new phenomenon is
  initially conceptualized and theorized. From this perspective, the digital
  sovereignty debate can be seen as an attempt to apply a classical concept to
  the digital realm. Here, the risk is that the attempt goes awry; sovereignty
  as a concept is stretched too much, such that it starts to lose its qualifying
  characteristics. The consequence would be a miscomparison. Although the four
concepts are simple and the coverage could be extended further, they retain
their powerfulness particularly for approaching theoretical concepts---and, if
anything, digital sovereignty is a still a vague and emerging theoretical
concept.

\section*{\change{The Peace of Westphalia as a Historical Parallel}}

The year 1648 haunts everyone participating in the current Internet governance
and digital sovereignty debate. But why \change{is something that happened 373
years ago relevant for the debate? In 1648 the} Peace of Westphalia was signed. It
ended the war that for thirty years had caused misery in Europe. Later on, much
later on, this peace treaty has been allegedly identified as the beginning of
the contemporary international system. Therefore, it also belongs to the
identity of international relations that study this system, the kernel of which
is based on the concept of sovereignty, and particularly the notion that each
state is sovereign within its own territory.

But sovereignty itself is a much older concept. From ancient Rome through the
medieval times to the Age of Enlightenment, sovereignty was essentially framed
as a question about the power of a ruler, the sovereign, over its
subordinates. Bodin, Hobbes, Locke, and Rousseau; from divinity to the concept
of popular sovereignty, a ruler who rules with the consent of people.  Here, a
dichotomous distinction between public and private is important: regardless
whether the legitimacy of the rule is based on divinity or consent, a sovereign
represents the public instead of the private
interests~\citep[p.~73]{Fukuyama15}. But there is more: ``\textit{on a given
  territory only one state can have sovereignty, that is, supreme authority, and
  that no other state has the right to perform governmental acts on its
  territory without its consent}'' \citep[p.~245]{Morgenthau49}. But with
respect to the digital sovereignty debate, is there a paradox, a treachery in
this image and where is it?

Clearly, another keyword here is territory. Actually, it can be argued that
territory, whether real or (and) imagined \change{by its occupants}---a
non-trivial distinction, is the only solid yet continuously moving boundary line
for sovereignty; without it there would not be a sovereign according to many
classical theorists~\change{(\citealt[pp.~71--74]{Lombardo15}; for the imaged
  part cf.~also \citealt{Anderson91})}. This argument opens the door for
conventional criticism about digital sovereignty: because the digital realm does
not follow geographic borders, there cannot be a supreme authority over the
realm~(cf.~\citealt{Mueller19}). While not fool-proof, this criticism has its
philosophical merits. But it is not where the image's treachery is located. Do
you see it already? Correct: it is in the quotation's second sentence claiming
that no other state can, without a consent, exert power in a territory of
another sovereign state.

The late 2010s digital ping pong across the Atlantic provides one of the means
with which the definition quoted can be criticized. The game played is also good
drama. Particularly jolly are the frequent episodes depicting the players
swinging in slow motion even though the ball has been taken away from them;
namely, by Schrems~(I) in 2015 and Schrems~(II) in 2020. The latter case
annulled the so-called Privacy Shield arrangement on transatlantic digital data
transfers, which had been established to replace the so-called Safe Harbor
arrangement that the former case had destroyed. In terms of sovereignty, the
former case is particularly relevant as the corresponding court decision
emphasized the supreme authority of national data protection authorities with
respect to the European Commission on one hand and with respect to the states
outside of the supranational union on the other. Thus, in essence, both the
legal and institutional bases were questionable for the Commission to use its
power to make the two decisions over transatlantic data flows (for a thorough
treatment of the current situation see \citealt{Christakis21}). But there is
more.

With respect to ping, extraterritorial power---the ability of a sovereign to
exert governmental actions in another sovereign's realm without its
consent---has long been a part of ping's data protection legislation. Already
before the General Data Protection Regulation (GDPR), the extraterritorial reach
was specified in Directive 95/46/EC, and contested in the famous Google Spain
case~\citep{Kindt16}. It was also the GDPR's sharpened extraterritoriality
provisions that provoked many critical comments from pong. That said, the same
year the GDPR came into force, pong passed its Clarifying Lawful Overseas Use of
Data Act (CLOUD Act). It provides pong's authorities access to data stored in
cloud services without the cumbersome mutual assistance treaty, effectively
deprecating ping's people constitutional protections provided by a jurisdiction
where the data is stored~\change{(see \citealt{Siry19}; and \citealt{Osula15}
  for an international law perspective). The same goes for the spicier cases;
  the mass surveillance programs with which a country bypasses another country's
  laws due to the borderless nature of the Internet. Snowden was merely a case;
  the show must go on, for both justifiable and unjustifiable
  reasons~\citep[cf.][]{BBC21a}.} From a theoretical point of view, these
examples exemplify how supreme authorities are trying to align themselves
\change{and their jurisdictions} with technology by using conceptual stretching,
\change{and bypassing international norms and laws along the way---as there is
  no way to conduct allegedly nearly global mass surveillance otherwise. As
  noted, however, every coin has two sides; transborder access to digital
  assets is also an important tool in criminal investigations.}

Sovereignty as a concept is ``\textit{plastic and evolving}'' \citep{Mueller20a}
to all directions; as it has always done, sovereignty transforms and
expands. However, the transformations through the stretching have clearly led to
paradoxes. Besides (a)~the misplaced authority of the Commission over the
decisions, which, in retrospect, can \change{be} interpreted as a theoretical
miscomparison regarding sovereign power, (b)~the CLOUD Act demonstrates that the
EU's complete territorial sovereignty over its citizens' personal data has been
legally challenged by the United States even without the two Schrems
decisions. Nor is the analogy of a ping-pong game by accident; (c)~the ongoing
sequence of events in the transatlantic data protection game demonstrates also
the strategic element present in the digital sovereignty affair. When one state
stretches sovereignty, other states respond with their own stretching,
shrinking, or something else. The data protection example further
(d)~illustrates how the paper's two themes---territorial sovereignty and human
autonomy---are theoretically connected. From this perspective, the EU's data
protection strategy ever since the 1990s can be also interpreted as an attempt
to provide autonomy for Europeans over their own data. If strategy as a concept
carries cynical connotations, as it often does in politics, a rephrasing is
easily available: the attempt can be seen as a milestone in the efforts to
establish human rights in the digital realm. But reaching such a milestone seems
to imply theoretically difficult and practically contested conceptual stretching
of territorial sovereignty.

\change{After this detour, it is worth returning to 1648 in which the Thirty
  Years' War ended. Three points are relevant for the digital sovereignty
  debate. First}, there is a debate about whether the peace treaty established
the system of international relations we know today. There are both proponents
\citep{Philpott99} and opponents~\citep{Osiander01, Sen12}. The debate has also
been seized to argue that there cannot be sovereignty in the digital
realm~\citep{Mueller19}. But are we seeing the same historical image? It was not
the only war. At the same time, the Eighty Years' War raged, having started from
Spain's attempt in the 1560s to crush the revolts in the Netherlands. The
longevity of these conflicts is astonishing for a today's reader. Indeed,
alongside military \textit{innovations}, the perhaps most striking element in
the conflicts was the \textit{capacity} of the states to engage in active
warfare, year after year and decade after decade, and to bear the burdens of the
wars, including the extreme human losses and financial costs. What the
Westphalian settlement brought was at least some religious and political
\textit{balance} across the Holy Roman Empire whose imperial power was limited
but not entirely eliminated. \citep[pp.~40--93.]{Kennedy88} Although the
settlement's relation to territorial sovereignty may be disputed, state capacity
and balance of power are still key concepts in international relations. So is
what threatens sovereignty, war.

\change{Second,} innovations and capacity, but not balance, characterize also
today's digital conflicts between states~\citep{Ruohonen20JCP}. Even though
different states have different stances on these conflicts just as they have
different positions on digital sovereignty~\citep{Chong14, Couture19,
  Mueller19}, what seems clear is that the line between war and peace is
becoming foggy in the digital realm, or that it has already become
blurry. Warfare in the digital realm may be a myth~\citep{Gartzke13}, or it may
be that obscurity prevents seeing the treachery of the image. But we can read,
and by reading we can learn that while a ``\textit{policy of ambiguity and
  silence}'' has been common, at least some states have agreed that sovereignty
applies in the digital realm together with Article~51 of the Charter of the
United Nations \citep{Moynihan19}. This article defines a right to self-defense
for a state or a collective of states in case of an armed attack. An armed
attack is a direct violation of a state's sovereignty in a territory. But it has
already been noted that the digital realm is not bound to territories, and,
hence, in theoretical terms, there cannot be armed conflicts in the realm. The
theoretical paradox is thus clear. For addressing this quagmire, recent
discussion has contemplated about thresholds for attacks in the digital
realm~\citep{Assaf20, ISS20a}. Against this backdrop, it is possible to continue
with another miscomparison involving the apparent presence of degreeism. What is
a ``90 percent peace'' or a ``10 percent war''?  Is parochialism further
present? Maybe it is not war \textit{or} peace but war \textit{and} peace?

\change{Third, the peace of Westphalia is important for understanding the
  historical developments behind the notion of a modern state in Germany and
  elsewhere. Before the peace, during the} long and rancorous politico-religious
war, which was partially fueled by the invention of the printing press that led
to the invention of modern propaganda, there were hundreds of sovereigns in
Germany, nominally united by the transnational Holy Roman Empire. None of the
sovereigns possessed the capacity of a state. They had no monopoly of force in
their territories. Their administrations were weak; they were even unable to
raise professional armies through systematic taxing of their
territories. Instead, their wars, were fought by mercenaries hired with borrowed
money, and when money ran short, the mercenaries turned to raiding and
looting. The Peace of Westphalia brought the impetus for a change. In a course
of about hundred years, the independent militias were disbanded, financial
control was established through a common bureaucracy, and military authority was
centralized. Army and bureaucracy were the
innovations.~(\citealt[pp.~67--70]{Fukuyama15}; \citealt{Graham11}.) In terms of
balance of power, the reduction in the number of European players eventually
followed~\citep[p.~270]{Morgenthau49}, as did the principle of non-intervention
into states' territorial sovereignty~\citep{Philpott99}. \change{In addition to
  the relevance for the history of international relations, the Holy Roman
  Empire provides a useful historical parallel; in some parts and to some
  extent, it can be used to reflect the state of the European Union as well as
  the state of the Internet (for the latter see \citealt{GurriRoberts20}).}  For
the present purposes, however, the lesson from Westphalia has more to do with
the capacity of a state to rule than with the territorial notion of
sovereignty. It is this lesson that the criticism of digital sovereignty tends
to miss---or, intentionally or unintentionally, obscure \change{due to
  parochialism of thinking}.

There \change{is also another lesson} closely related to Westphalia. When
fast-forwarding in time, it is possible to find another kind of a
  sovereign via conceptual stretching. That is, a private company raging
private or quasi-private wars to advance commercial goals; and, again, military
innovation and the capacity to exert power, but not balance, were present just
as sovereignty was~\citep{Robins02, Sen12}. While the East India Company was
eventually dissolved, certain parallels are present with our times. In someone
else's words:

\begin{quote}
``\textit{Sadder still to think that if this is a new realm of national sovereignty then our existing nation-state world order is just simply not able to engage with the new IT corporate nation-states in any manner that can curb their overarching power to defend their chosen borders. The 1648 Peace of Westphalia has much to teach us, and not all of the lesson is pleasant.}'' \citep{Huston20a}
\end{quote}

As is often the case with history, a sense of irony is present: states and
international organizations greatly contributed to the expansion of the digital
realm throughout the 20th century~\citep[pp.~571--599]{Cortada12}, but they,
willingly or unwillingly, lost their control during the early 21th
century. Again via conceptual stretching, it is possible to argue that a new
form of sovereignty emerged in the realm, a ``\textit{de facto digital corporate
  sovereignty}''~\citep{Floridi20}. Unlike the East India Company, the new
digital corporate sovereignty is not geographically bound, which is again
paradoxical because states and the legitimacy for their power are still largely
dependent on territories. But now, in 2021, it seems undisputed that states are
nevertheless determined to fight back, partially due to the many unforeseen
consequences from the loss of control, and partially due to other states who
retained their control. The task is easier said than done. But before briefly
contemplating about obstacles, the other side in the debate, the human-centered
autonomy, should be tackled. As the task is formidable, in what follows, the
broader digital sovereignty debate again frames the discussion and deliberation.

\section*{Toward Human Autonomy in the Digital Realm}

For lack of a better term, let ``digital self-sovereignty'' describe
the autonomy of individuals in the digital realm. Think of it as a superset to
which different sover\-eignty-related subsets belong. The most important subset
is the autonomy of individuals vis-\`a-vis a sovereign, the supreme
authority---or authorities if one adopts the viewpoint that there are no
territorial sovereignties in the digital realm.

Here, again, we are confronted with centuries of thought. For Hobbes, writing
three years after the Peace of Westphalia, during the English Civil War, an
absolute sovereign was needed because otherwise everyone would be in a war
against each other. For later theoretical extremists such as Schmitt, in
  particular, it was precisely this state of anarchy, where everyone is a
  \textit{lupus}, a wolf, to one another, that revealed the true essence of
  politics, which, according to Schmitt, cannot be tamed by
  liberalism~\citep{Sartori89}. This philosophical anarchy was also later
hijacked to describe the international relations between states, and, by
extension, state actors in the digital realm, although the pirating was not
entirely accurate because for Hobbes the anarchy was between individuals and not
states~\citep{Christov17, Ruohonen20JCP}. For this reason and other reasons, we
must look elsewhere to attach our thought. To do so, we again fast-forward in
time.

In the early 19th century, Constant, during the Napoleonic Wars and their
terror, published a set of principles for ``all governments''. By attacking
Hobbes directly, he advocated popular sovereignty but was eager to limit its
power against individuals---as power is evil and dangerous, it was to be
constrained by institutions and individual freedoms, among them the freedom of
thought, as it was also language through which despotism
triumphed~(\citealt{Garsten12}; \citealt[pp.~65--67]{Rosenblatt18}). From this,
we can crystallize the first subset of the superset as the rule of law, which,
among other things, limits politics and power by law. Then, individuals can pool
``\textit{their self-sovereignties (sovereignty on themselves) through
  deliberation, negotiation, and voting, to create popular sovereignty, which
  then legitimises national sovereignty, which then controls individuals' legal
  exercise of their self-sovereignties}''~\citep[p.~375]{Floridi20}. But as
beautiful as this idea is, and despite the optimism expressed by
some~\citep{Noveck20}, the deliberation required for digital self-sovereignty
can hardly be argued to work properly. Hence, also the creation of a popular
sovereignty remains incomplete, which then transforms into a partial
delegitimization of a national sovereignty, which then transforms to a partial
deprecation of individuals' self-sovereignties.

Why is \change{the digital self-sovereignty} not working? To answer to the
question, we need to look for other subsets, for other freedoms Constant and his
later followers put forward. From the Universal Declaration of Human Rights we
can find many of the necessities for digital self-sovereignty, among them the
rule of law provisions, the freedom of thought, and the right to privacy and
honor. These can be further extended, as the EU has done with its right to data
protection, the basis of the GDPR, for instance. Thus, all the ingredients are
there---international law, human rights, and so forth; yet it is not
working. \change{To seek for a partial explanation, the Thirty Years' War again
  provides a useful historical parallel;} there were bandits pillaging, wars
that seemed eternal, and \change{sovereigns} who were not states. \change{Thus,}
the analogy with the current digital realm \change{is} clear, whether the
context is cyber warfare or the ongoing online information
disorder~\citep[cf.][]{Ruohonen20JCP, Ruohonen21GOODIT}.  Earlier, we also noted
the rivalry between ping and pong. The same rivalry applies to privacy
legislations, although their positions are not necessarily too far
apart~\citep{Buyuksagis19, Christakis21}. From ping's experience we can also
learn how difficult it is to enforce the rule of law in the digital realm, at
least at short notice, as the GDPR's early enforcement problems
demonstrate~\citep{Ruohonen20IS}. From the lesson of Westphalia we learned that
it took almost a hundred years or so for a change to occur. Indeed, not all of
the lesson is pleasant. Maybe it is Hobbes instead of Constant after all;
\change{conflicts and anarchy instead of liberalism and rights}?

Are there more paradoxes? The functioning of the Internet is often described
with the so-called Open Systems Interconnection (OSI) model (see, e.g.,
\citealt{KuroseRoss08}). Although it has never entirely accurately described the
reality, it is a useful analytical framework akin to the concept of ideal types
in social sciences. The model has seven layers; the lower a layer is located,
the more low-level the technical details. As said, much of the action visible to
end-users occurs at the seventh layer, at the application layer. It is also this
layer through which individuals express their digital self-sovereignties. But it
is the third layer, the network layer, that is fundamental to the functioning of
the Internet. Regarding this layer, some have recently expressed fears that
changes, such as regulation, at the application layer will spill over to the
network layer, or that new protocols will be rolled out for this layer,
potentially causing unintended changes and fragmentation~\citep{Ohara20,
  Mueller20a}. While such fears---rational or irrational as they may
be---reflect the critical side in the digital sovereignty debate, there should
be no reason to shy away from also acknowledging the current problems, such as
well-known issues with routing~\citep{Hesselman20, Testart19}. To some degree,
conceptual shrinking is present; the debate is directed toward narrow technical
topics.

In the big picture, in contrast\change{---and besides the mass surveillance
  programs supposedly operating already at the network layer or below, } it is
increasingly difficult to separate different layers due to the East India
Companies who provide everything from fiber to applications. Centralization
follows~\citep[cf.][]{Moura20}. By implication, it is not clear whether the OSI
model is applicable for the digital sovereignty debate.  In addition to
companies' technical strong-arm over sovereignty, they also control and govern,
without the rule of law, the rights and freedoms, and, thus, individuals'
self-sovereignties, which provide the popular sovereignties, which transform to
national (territorial) sovereignties.

Data is a further subset in the superset. Ever since the 1983 decision of the
German constitutional court, the European perspective has relied on information
self-determination; personal data belongs to a person from whom it was
extracted. \change{Yet,} in terms of self-sovereignty in the digital realm, the
bargain has long been about ``\textit{selling your soul while negotiating the
  conditions}'' \citep{Belli17}, a phrasing that refers to the dubiousness of
consenting to the conditions of digital services and applications. For this
reason, the GDPR, like other emerging legislations~\citep{Tunney20}, have moved,
or at least have tried to move, away from consent as \change{a single} mechanism
for the rule of law in this context. At the same time, in Europe, there has been
a trend toward strengthening self-sovereignty through voluntary data-sharing
that benefits also the economy.

Here, too, the image suffers from a paradox; do they understand, when they sell
their souls \change{for voluntary data sharing}? A thorough literature search of
empirical studies is not necessary to put forward \change{an} answer that most
of them do not understand what they are consenting to (for the background
\change{of this privacy paradox} see, e.g., \citealt{Norberg07}). There are no
reasons to suspect that this claim would be different for altruistic
data-sharing. While the goals are noble, masquerading as empowerment and agency,
there has thus been a tension between those who portray self-sovereignty through
consumerism and those who value citizenship \change{and
  rights}~\citep{Lehtiniemi20}. Further issues are easy to
pinpoint. Particularly in the industrial setting, data sovereignty has been
understood to mean that you own and can control your own data; you own and can
control the data that your machinery in your factory continuously
produces~\citep{Jarke20}. But there are also employees in your factory who
increasingly generate data on their own when working with your machinery in your
factory. Who owns such data related to the means of production? \change{Answers
  to the question remain open.}

\change{There is also a} red flag in the digital sovereignty debate, data
localization. While it lingered in European policy circles throughout the 2010s,
often in disguise~\citep{Celeste20}, recent large-scale projects, such
\text{GAIA-X}, the pan-European cloud computing framework, have poured gasoline
to the fires kept by those critical to data localization. In general, the
technical details \citep[cf.][]{Hesselman20} are still too vague to make any
definite conclusions; therefore, it suffices to merely point out the critical
arguments that data localization in itself does not guarantee self-sovereignty
in terms of security and privacy of personal data~\citep{Komaitis17}. Regardless
whichever side one takes in this debate, the theoretical underpinning is
essentially again about the intervening and alignment of the two
realms. \change{Finally, it is worth noting a point that national and
  transnational digital infrastructure projects can also be portrayed as an
  attempt to build imagined national (European) digital identities in the new
  digital era~\citep{Mollers21}. Here, again, the Holy Roman Empire, the peace
  of Westphalia, the later age of great powers, and the East India Company
  provide good historical parallels because nationalism and national identities
  emerged much later in many European countries. By and large, although not
  everywhere, the modern nation-states and their supranational unions were also
  forged on anvils of war.}

\section*{Reflections}

This short theoretical essay asked us to think and participate in the debate on
digital sovereignty. It urged us to reflect upon the treacheries of images,
including the implicit political dogmas and academic miscomparisons. We did
so. But what did we see? Together, hopefully, we saw some important reflections,
and beyond. To briefly summarize the paradoxes arising from different
miscomparisons, it is clear that classical definitions of state sovereignty,
such as the one provided by \citet{Morgenthau49}, align only poorly with the
digital realm. The GDPR and the CLOUD Act together provide a good example in
this regard; theoretically, these can be seen to stretch and shrink,
respectively, the territorial notion of sovereignty. Stretching of the concept
seems also necessary in order to achieve digital sovereignty for the
subordinates of a democratic state sovereign. But such stretching is difficult
even among democracies due to international relations that are still largely
based on Westphalian and later ideas about territorial boundaries. Given these
reasons and the paradoxes elaborated, strategic autonomy seems a better concept
theoretically.

The same applies to state power, which, too, at least thus far, is ultimately
territorial with respect to both other states and the subordinates within
states. Here lies the crux of the paper's contribution to the digital
sovereignty debate. For the early utopian theorists, \change{such as
  \citet{Barlow96}}, the cyber space was a sovereign space by itself, but since
there was no authority, no Leviathan, which is required for sovereignty as a
theoretical construct without losing its qualifying characteristics, the
discourse was later hijacked by ``realists'' who perceived the cyber space as an
anarchy akin to Hobbes, Schmitt, and other conflict
theorists~\citep[cf.][]{Mueller19}. The fundamental theoretical paradox is that
much of Western philosophical thought has been either intentionally or
inexpertly skipped with this utopian-realist parochialism. Among these omissions
is liberalism \change{together with international law and norms}. If there is
only a sovereign space without a sovereign authority, can there be freedom,
liberty, human rights, democracy, etc., and can these even survive in such a
space? The peace of Westphalia has three painful lessons for a deliberation
about this question.

The first is administration---and not ``governance''. To rule---to move from the
``code is law'' idiom to a ``law is code'' notion \citep{Timmers19a}, is to have
a well-functioning administration, a bureaucracy that ultimately enforces the
Weber's definition for the monopoly of force. (To avoid dogmas, it should be
added that no one participating in the debate has advocated anything like
Weber's iron cage of bureaucracy.) In addition, the collective decisions made
must be legitimate and overrule all other decisions, and they cannot provide an
exit in a sense that they would not cover a whole territory and its
population. These are lessons from the Peace of Westphalia. With these lessons
the concept of sovereignty, as well as the concept of politics, is expanded to
something different than what territory and morality can provide. When taken to
the theoretical extreme, the concept of sovereignty transforms into a sovereign
as an exception~\citep{Sartori89}, and it has always been the rule of law that
constraints sovereignty as an exception, as well as the Hobbesian anarchy. If
anything, the lessons from Westphalia are also lessons about warfare. Therefore,
territorial sovereignty cannot be ignored. Unfortunately, there are no easy
answers to a question about how the digital realm, law, and territory could be
combined; one option would be to extend sovereignty to cover administration,
such that interference with a state's core decision-making processes would
violate law~\citep{Midson14}. But what is a core decision-making process? In a
democracy digital deliberation is, and must be, open and public, and, therefore,
it is also vulnerable to new threats such as disinformation. Even though there
are again no easy answers on the horizon, our reflection pointed toward digital
self-sovereignties of individuals, which, in a democracy, are directly related
to popular sovereignty, which is related to administration, which is related to
other things. In this difficult puzzle the different pieces reinforce each
other.

But the fundamental problem is that the pieces are not reinforced together in
the puzzle. In the big picture, there are three problems. None of these have
easy answers. First, the technical Internet community, including standardization
organizations, committees, and talkshops, is unlikely to be able to solve the
problems alone~\citep{Ohara20}. Nor is it clear whether the community's views
reflect the views of the self-sovereigns in the Internet. Aligning the community
with the global civil society community might be an option, but the latter seems
to be too fragmented, lacking sufficient political legitimacy and thus
power. Further alignment is required. But this alignment brings the second
problem: the international relations, power politics, and, more or less,
anarchy. In this context it seems again appropriate to quote someone else:

\begin{quote}
``\textit{International law is a decentralized legal order in a dual sense. In the first place, its rules are, as a matter of principle, binding only upon those states which have consented to them. In the second place, many of the rules that are binding by virtue of the consent given are so vague and ambiguous and so qualified by conditions and reservations as to allow the individual states a very great degree of freedom of action whenever they are called upon to comply with a rule of international law}.'' \citep[p.~244]{Morgenthau49}
\end{quote}

This quotation summarizes many points in our reflection. There is again the
concept of consent. Not everyone consents, whether they are states or
self-sove\-reign individuals and East India Companies. Then there are the leeway
and ambiguity so typical to international arrangements, whether laws or
norms. Particularly in the digital realm, the international arrangements are
often a bizarre ``\textit{shadow theater}'' carefully constructed by those
states who benefit from the \textit{status~quo}~\citep{Tikk18}, or want to
benefit from it, or simply do not care. When stretched a little, the quotation
can be read also as a statement about enforcing laws in the digital realm. If
the word international is replaced with the word regional, the quotation could
also describe some laws enacted in the European Union. \change{Despite rigorous
  legal safeguards and frameworks established in Europe, including the GDPR and
  associated laws, state surveillance has continued alongside commercial
  surveillance. While controlling the latter through the power granted by the
  EU's internal market is difficult but still possible, the former seems to be
  generally beyond the power of the EU and its institutions. Irrespective of
  several high-profile cases fought and won in the highest European courts,
  national security concerns have kept surveillance capabilities firmly on the
  table. A~good example would be data localization and its relation to data
  retention laws enacted---and systematically violated---throughout Europe (for
  the background see \citealt{Albrecht20}; \citealt{GrabowskaMoroz20}). Thus, to
  some extent, a miscomparison is present; the degree of freedom for states to
  comply with international law applies also to the EU and its member states.}
Not all laws are equal and not all laws are enforced with the same vigor; here
we have the eternal lesson about law-making---laws that are not laws are mere
norms.

The third and final problem is the most important. It is about functioning of
political systems. Without functioning systems, whether democratic or something
else, it is impossible to solve the previous two problems, provided that there
is political will to solve them. At minimum, a functioning democratic political
system requires a well-functioning tripod: the state, the rule of law, and
political accountability, each reinforcing each
other~\citep[pp.~532--534]{Fukuyama15}. Today, technology should be perhaps
added to the rack as a fourth item, a well-functioning technology, that is. By
implication, the urge, if not pressure, for the world's tripods to work with
technology-makers is increasing day by day~\citep{Schneier20}. But there is
more. Political accountability implies that the output from a political system
is legitimate. For an output to be legitimate, there must be also input to a
democratic political system. The input cannot be from technology-makers
alone. Nor cannot it occur only through elections. Deliberation is required by
informed citizens. For deliberation to work, the voice of everyone affected by a
particular policy should be heard; however, the input only qualifies as valid
input when it is backed by qualified reason rather than pure self-interest,
force, or manipulation~\citep{Gaus20}. Deliberation further requires public
channels in order to deliver inputs to sovereigns who represent the public
interests. Disinformation and related information warfare tactics can
effectively destroy such delivery channels. For these and related reasons, the
self-sovereignty umbrella concept sketched should not be taken lightly---even if
further theorization is required to sharpen the concept.

\section*{\change{Conclusion}}

\change{This theoretical essay addressed digital sovereignty by focusing on its
  territorial and human-centered aspects. The primary conclusion is clear and
  simple: as a theoretical concept, sovereignty aligns poorly with the digital
  realm, including the Internet. The main reason is the concept's close ties to
  territoriality and geographic boundaries of states, their supranational
  unions, and their alliances. In many respects, international law and related
  aspects rely on the same boundaries. This reliance on geography makes it also
  difficult to achieve human-centered autonomy in the digital realm; because
  there is no supreme authority in the realm, it is difficult to enact and
  particularly enforce laws in the realm.}

\change{Attempts to overcome the difficulty has resulted different theoretical
  paradoxes, whether in terms of surveillance, transborder access to digital
  assets, or recent data protection laws, which all tend to contradict the
  classical territorial notion of state sovereignty. Several historical
  parallels can be used to illustrate these paradoxes. Among these are the
  famous peace of Westphalia and the Holy Roman Empire. In addition, the equally
  famous East India Company provides an interesting albeit not perfect
  historical parallel for the concept of digital corporate sovereignty. Further
  paradoxes emerge with the human-centered autonomy: the famous privacy paradox,
  the ownership of data, the vagueness of consent for data protection and
  privacy, surveillance, and the complex relation between data localization,
  data retention, and digital self-sovereignty. Through the historical
  parallels, the argument put forward is simple: to move forward, to at least
  partially address the paradoxes and their practical political ramifications,
  efficient administration is necessary alongside finding and theorizing new
  openings that go beyond the simplification of realism and
  utopianism. Pioneering new ideas is arguably also necessary to advance
  democracy in the new digital era.}

\section*{Acknowledgements}

This \change{theoretical} essay was funded by the Strategic Research Council at the Academy of Finland (grant number~327391). The author thanks Paul Timmers for an earlier exchange of ideas.

\bibliographystyle{apalike}

\end{document}